\documentclass{iopart}  
\usepackage{epsfig,subfigure}
\usepackage{latexsym}
\begin{document}

\letter{A cellular-automata model of flow in ant-trails:\\
non-monotonic variation of speed with density}
\author{Debashish Chowdhury$^{1,2}${\footnote{E-mail:debch@iitk.ac.in}}, 
Vishwesha Guttal$^1$, 
Katsuhiro Nishinari$^{2,3}${\footnote{E-mail:knishi@rins.ryukoku.ac.jp}} 
and Andreas Schadschneider$^2${\footnote{E-mail:as@thp.uni-koeln.de}}}
\address{$^1$\ Department of Physics, Indian Institute of Technology, 
Kanpur 208016, India.}
\address{$^2$\ Institute for Theoretical Physics, University of Cologne, 
50923 K\"oln, Germany.}
\address{$^3$\ Department of Applied Mathematics and Informatics, 
Ryukoku University, Shiga 520-2194, Japan.}

\vspace{0.7cm}
%\submitted
\begin{center}
published in {J.\ Phys.\ A35, L573 (2002)}
\end{center}
%\date{}

\begin{abstract}
Generically, in models of driven interacting particles the 
average speed of the particles decreases monotonically with 
increasing density. We propose a counter-example, motivated by the 
motion of ants in a trail, where the average speed of the particles 
varies {\it non-monotonically} with their density because of the 
coupling of their dynamics with another dynamical variable.
These results, in principle, can be tested experimentally. 
\end{abstract}

Particle-hopping models, formulated usually in terms of cellular 
automata (CA) \cite{wolfram}, have been used to study the 
spatio-temporal organization in systems of interacting particles 
driven far from equilibrium \cite{sz,schutz,md,droz} which include, 
for example, vehicular traffic \cite{css,helbing}. In general, 
the inter-particle interactions tend to hinder their motions so 
that the average speed decreases {\it monotonically} with the 
increasing density of the particles. In this letter we report a 
counter-example, motivated by the motion of ants in a trail 
\cite{burd}, where the average speed of the particles varies 
{\it non-monotonically} with their density because of the coupling 
of their dynamics with another dynamical variable.

The ants communicate with each other through a process called 
chemotaxis by dropping a chemical (generically called {\it pheromone}) 
on the substrate as they crawl forward \cite{wilson,camazine}. 
Although we cannot smell it, the trail 
pheromone sticks to the substrate long enough for the other following 
sniffing ants to pick up its smell and follow the trail. In this 
letter we develope a CA model which may be interpreted as a model of 
unidirectional flow in an ant-trail. Rather than addressing the 
question of the emergence of the ant-trail, we focus attention here 
on the traffic of ants on a trail which has already been formed.

\begin{figure}[h]
\begin{center}
\includegraphics[width=0.8\columnwidth]{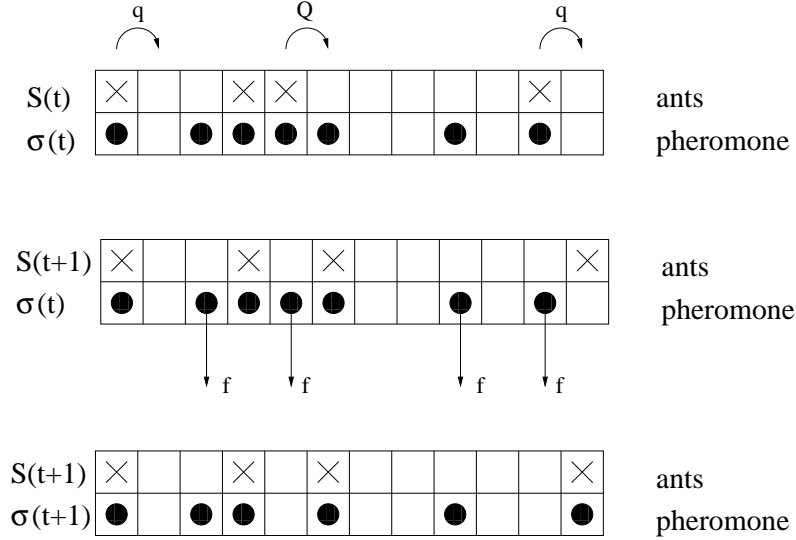}
\end{center}
\caption{ Schematic representation of typical configurations; it 
also illustrates the update procedure. Top: Configuration at time $t$, 
i.e.\ {\it before} stage $I$ of the update. The non-vanishing hopping 
probabilities of the ants are also shown explicitly. Middle: 
Configuration {\it after} one possible realisation of {\it stage $I$}. 
Two ants have moved compared to the top part of the figure. Also 
indicated are the pheromones that may evaporate in stage $II$ of the 
update scheme.  Bottom: Configuration {\it after} one possible realization 
of {\it stage $II$}. Two pheromones have evaporated and one pheromone has 
been created due to the motion of an ant.
 }
\label{fig-0}
\end{figure}

Each site of our one-dimensional ant-trail model represents a cell 
that can accomodate at most one ant at a time (see fig.\ref{fig-0}). 
The lattice sites are labelled by the index $i$ ($i = 1,2,...,L$); 
$L$ being the length of the lattice. We associate two binary variables 
$S_i$ and $\sigma_i$ with each site $i$ where $S_i$ takes the value 
$0$ or $1$ depending on whether the cell is empty or occupied by an ant. 
Similarly, $\sigma_i =  1$ if the cell $i$ contains pheromone; otherwise, 
$\sigma_i =  0$. Thus, we have two subsets of dynamical variables in 
this model, namely, 
$\{S(t)\} \equiv (S_1(t),S_2(t),...,S_i(t),...,S_L(t))$ 
and 
$\{\sigma(t)\} \equiv (\sigma_1(t),\sigma_2(t),...,\sigma_i(t),...,
\sigma_L(t))$.
The instantaneous state (i.e., the configuration) of the system at 
any time is specified completely by the set $(\{S\},\{\sigma\})$.

We assume that the ant does not move backward; its forward-hopping 
probability, however, is higher if it smells pheromone ahead of it. 
The state of the system is updated at each time step in {\it two 
stages}.  At the end of stage I we obtain the subset 
$\{S(t+1)\}$ at the time step $t+1$ using the full information 
$(\{S(t)\},\{\sigma(t)\})$ at time $t$. At the end of the stage II 
we obtain the subset $\{\sigma(t+1)\}$ at the time step $t+1$ using 
the subsets $\{S(t+1)\}$ and $\{\sigma(t)\}$.\\

\noindent {\it Stage I}: The subset $\{S\}$ (i.e., the {\it positions} 
of the ants) is updated {\it in parallel} according to the following 
rules:\\[0.2cm]
\noindent If $S_i(t) = 1$, i.e., the cell $i$ is occupied by an ant at the 
time step $t$, then the ant hops forward to the next cell $i+1$ with\\
\begin{equation}
{\rm probability} = \left\{\begin{array}{lll}
            Q &~{\rm if\ }~ S_{i+1}(t) = 0 ~{\rm but}~ \sigma_{i+1}(t) = 1,\\  
            q &~{\rm if\ }~ S_{i+1}(t) = 0 ~{\rm and}~ \sigma_{i+1}(t) = 0,\\ 
            0 &~{\rm if\ }~ S_{i+1}(t) = 1. 
\end{array} \right.
\end{equation}
where, to be consistent with real ant-trails, we assume $ q < Q$.\\

\noindent {\it Stage II}: The subset $\{\sigma\}$ (i.e., the presence or 
absence of pheromones) is updated {\it in parallel} according to the 
following rules:\\     

\noindent If $\sigma_i(t) = 1$, i.e., the cell $i$ contains pheromone at 
the time step $t$, then it contains pheromone also in the next time step, 
i.e., $\sigma_i(t+1) = 1$, with \\
\begin{equation}
{\rm probability} = \left\{\begin{array}{lll}
            1 &~{\rm if}~ S_i(t+1) = 1 ~{\rm at ~the ~end ~of ~stage ~I,}\\
          1-f &~{\rm if}~ S_i(t+1) = 0 ~{\rm at ~the ~end ~of ~stage ~I}. 
\end{array} \right.
\end{equation}
where $f$ is the pheromone evaporation probability per unit time.

\noindent On the other hand, if $\sigma_i(t) = 0$, i.e., the cell $i$ does not 
contain pheromone at the time step $t$, then 
\begin{equation}
\sigma_i(t+1) = 1 \ \ \ {\rm iff}~ S_i(t+1) = 1
{\rm ~at ~the ~end ~of ~stage ~I}.
\end{equation}

In certain limits our model reduces to the Nagel-Schreckenberg (NS) model 
{\footnote{By the term 'NS model'
in this letter we shall always mean the NS model with maximum allowed
speed {\it unity}, so that each particle can move forward, by one
lattice spacing, with probability $q_{NS}$ if the lattice site
immediately in front is empty.} \cite{ns} which is the minimal
particle-hopping model for vehicular traffic on freeways.
The most important quantity of interest in the context of flow 
properties of the traffic models is the {\it fundamental diagram}, 
i.e., the flux-versus-density relation, where flux is the product of 
the density and the average speed. For a hopping probability $q_{NS}$ 
at a given density $c$ the exact flux $F(c)$ in the NS model is given 
by \cite{css,ssni}
\begin{equation}
F_{NS}(c)= \frac{1}{2}\left[1-\sqrt{1 ~- ~4 ~q_{NS} ~c(1-c)}\right].
\end{equation}
which reduces to $F_{NS}(c) = \min(c,1-c)$ in the deterministic limit 
$q_{NS} = 1$.

Note that in the two special cases $f = 0$ and $f = 1$ the ant-trail 
model becomes identical to the NS model with $q_{NS} = Q$ and 
$q_{NS} = q$, respectively. Extensions of the NS model have been used 
not only to capture different aspects of vehicular traffic \cite{css} 
but also to simulate pedestrian dynamics \cite{helbing,helbing2,schad}. 
In a closely related CA model for pedestrian dynamics \cite{schad} 
the floor fields, albeit {\it virtual}, are analogs of 
the pheromone fields $\{\sigma\}$ in the ant-trail model. 
However, in the pedestrian model there is no exclusion principle for
the floor field.

\begin{figure}[h]
\begin{center}
\includegraphics[width=0.6\columnwidth]{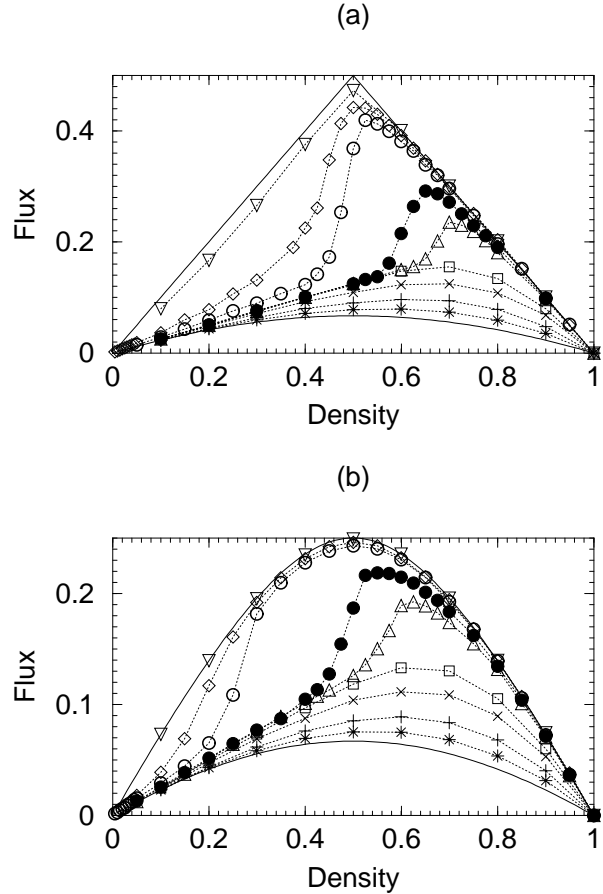}
\end{center}
\caption{ The flux of ants plotted against their densities for the 
parameters (a) $Q = 1, q = 0.25$ and (b) $Q = 0.75, q = 0.25$. 
The discrete data points corresponding to 
$f = 0.0001 ({\bigtriangledown})$, $0.0005 ({\Diamond})$, 
$0.001 (\circ$), $0.005 (\bullet)$, $0.01 ({\bigtriangleup})$, 
$0.05 ({\Box})$, $0.10 (\times)$, $0.25 (+)$, $0.50 (\ast)$ have been 
obtained from computer simulations; the dotted lines connecting these 
data points merely serve as the guide to the eye. The two continuous 
solid curves at the top and bottom correspond to the flux in the NS 
model for $q_{NS} = 1.0$ and $q_{NS} = 0.25$, respectively, in (a) 
and for $q_{NS} = 0.75$ and $q_{NS} = 0.25$, respectively, in (b). 
}
\label{fig-1}
\end{figure}

The ant-trail model we propose here is also closely related to the bus 
route model (BRM) \cite{loan,cd}. The variables $S$ and $\sigma$ in the 
ant-trail model are the analogs of the variables representing the 
presence (or absence) of bus and passengers, respectively, in the BRM. 
Because of the periodic boundary conditions, the number of ants and 
buses are conserved while the pheromone and passengers are not conserved. 
However, unlike the BRM, the pheromones are not dropped independently 
from outside, but by the ants themselves. Another crucial difference 
between these two models is that in the bus-route model $Q < q$ (as the 
buses must {\it slow down} to pick up the waiting passengers) whereas in 
our ant-trail model $Q > q$ (because an ant is more likely to move forward 
if it smells pheromone ahead of it). 

\begin{figure}[h]
\begin{center}
\includegraphics[width=0.60\columnwidth]{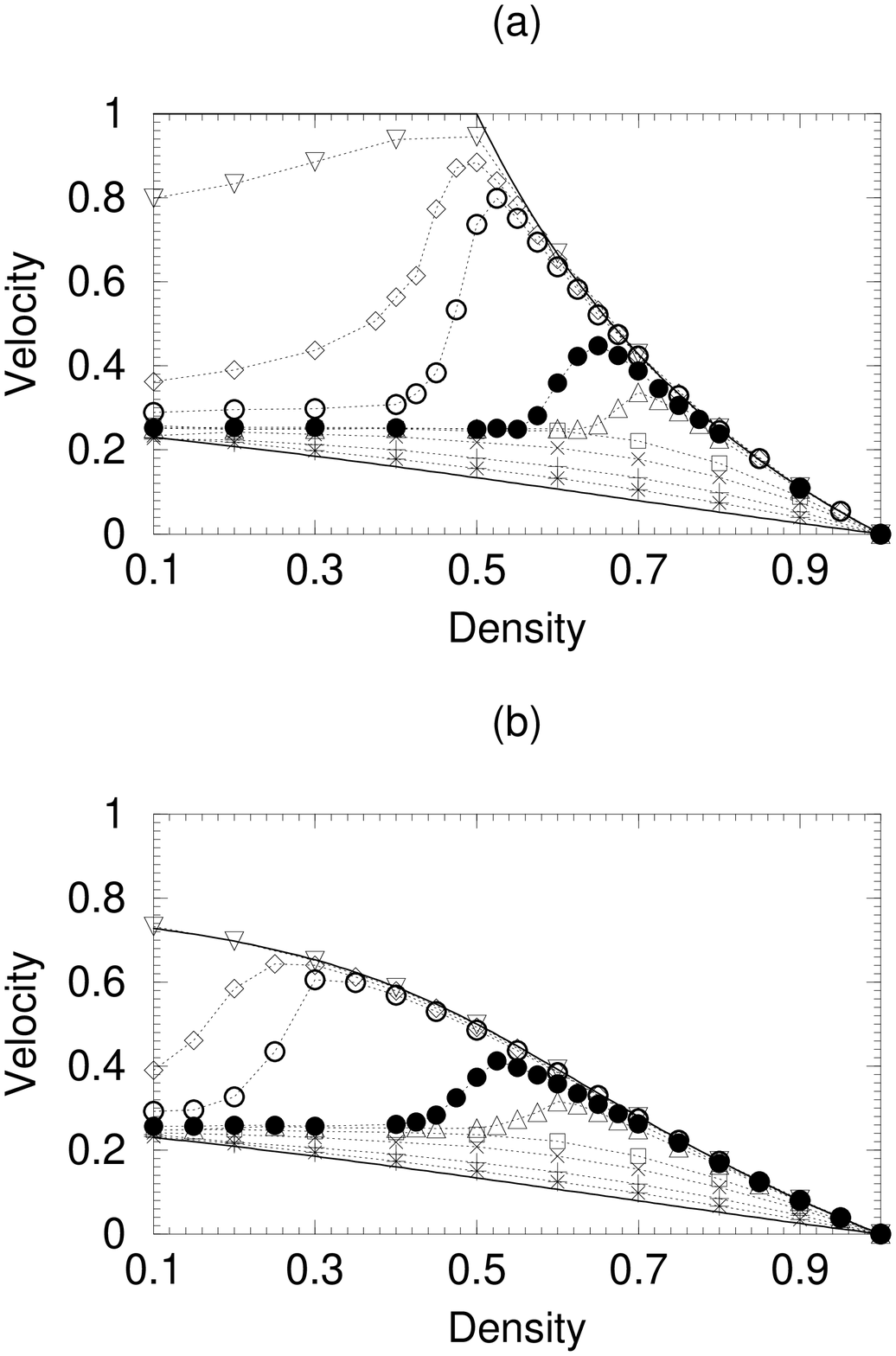}
\end{center}
\caption{The average velocity of ants plotted against their densities 
for the parameters (a) $Q = 1, q = 0.25$ and (b) $Q = 0.75, q = 0.25$. 
Same symbols in the figs.\ref{fig-1} and \ref{fig-2} correspond to the 
same values of the parameter $f$.} 
\label{fig-2}
\end{figure}

In fig.\ref{fig-1} we show the fundamental diagrams obtained by extensive 
computer simulations of the ant-trail model for several values of $f$. 
The most unusual features of the fundamental diagrams shown in 
fig.\ref{fig-1}  are that, over an intermediate range of values of 
$f$ (for example, $f = 0.0005, 0.001, 0.005, 0.01$ in fig.\ref{fig-1}) 
the flux in the low-density limit $c \rightarrow 0$ is very close to 
that for the NS model with $q_{NS} = q$  whereas in the high-density 
limit $c \rightarrow 1$  the flux for the same $f$ is almost identical 
to that for the NS model with $q_{NS} = Q$. 
These unusual features of the fundamental diagrams arise from the 
{\it non-monotonic} variation of the average velocity with the density 
of the ants (see fig.\ref{fig-2}).

The presence of the pheromone essentially introduces an {\it effective 
hopping probability} $q_{\rm {eff}}(c)$, which depends on the ant 
density $c$. The particle-hole symmetry (and hence the 
symmetry of the fundamental diagram about $c = 1/2$) observed in the 
special limits $f = 0$ and $f = 1$, are broken by the $c$-dependent 
effective hopping probability for all $ 0 < f < 1$ leading to a peak 
at $c > 1/2$. 
Furthermore, the analysis of correlation functions reveals some interesting
clustering properties which will be studied in detail in a future
publication \cite{chowdhury}.

The qualitative features of the $c$-dependence of $q_{\rm eff}$ can be
reproduced by an analytical argument based on a {\it mean-field 
approximation} (MFA) \cite{chowdhury}. In this MFA, let us assume 
that all the ants move with the mean velocity $\langle V\rangle$ which depends 
on the density $c$ of the ants as well as on $f$; although, to begin 
with, the nature of these dependences are not known we'll obtain 
these self-consistently.

Let us consider a pair of ants having a gap of $n$ sites in between.
The probability that the site immediately in front of the following 
ant contains pheromone is $(1-f)^{n/\langle V \rangle}$ since 
$\frac{n}{\langle V \rangle}$ is the average time since the pheromone
has been dropped. 
Therefore, in the MFA the effective hopping probability is given by 
\begin{equation}
q_{\rm eff} = Q (1-f)^{n/\langle V\rangle} + q \{1-(1-f)^{n/\langle 
V\rangle}\}.
\end{equation}
We replace $n$ by the corresponding exact global mean 
separation $\langle n\rangle = \frac{1}{c} - 1$ between successive ants. 
Moreover, since $\langle V\rangle$ is identical to $q_{\rm eff}$, we 
get the equation
\begin{equation}
\biggl(\frac{q_{\rm eff}-q}{Q-q}\biggr)^{q_{\rm eff}} = 
(1-f)^{\frac{1}{c}-1}
\label{eq-sc}
\end{equation}
which is to be solved self-consistently for $q_{\rm eff}$ as
a function of $c$ for a given $f$. Note that the equation (\ref{eq-sc})
implies that, for {\it given} $f$, $\lim_{c \rightarrow 0}q_{\rm eff} = q$; 
this reflects the fact that, in the low-density regime, the pheromone 
dropped by an ant gets enough time to completely evaporate before the 
following ant comes close enough to smell it. Equation (\ref{eq-sc})
also implies $\lim_{c \rightarrow 1} q_{\rm eff} = Q$; this captures 
the sufficiently high density situations where the ants are too close 
to miss the smell of the pheromone dropped by the leading ant unless 
the pheromone evaporation probability $f$ is very high. Similarly, 
from (\ref{eq-sc}) we get, for {\it given} $c$,
$\lim_{f \rightarrow 1} q_{\rm eff} = q$ and $\lim_{f \rightarrow 0} 
q_{\rm eff} = Q$
which are also consistent with intuitive expectations.

In view of the fact \cite{camazine} that the lifetime of pheromones 
can be as long as thirty to sixty minutes, the interesting regime 
of $f$ ($\ll 1$), where the average velocity varies non-monotonically 
with the ant density, seems to be experimentally accessible.  We hope 
that the non-trivial predictions of this minimal model of ant-trail 
will stimulate experimental measurement of the ant flux as a function 
of the ant density for different rates of pheromone evaporation by 
using different varieties of ants \cite{burd}.

\section*{Acknowledgment} We thank B. H\"olldobler for drawing our 
attention to the reference \cite{burd}.

\end{document}